\documentclass[prl,superscriptaddress, preprintnumbers,twocolumn,showpacs,amsmath,amssymb]{revtex4}
\usepackage{amsmath}
\usepackage{amssymb}
\usepackage{amsthm}
\usepackage{amsfonts}
\usepackage{algorithmic}
\usepackage{enumerate}
\usepackage{latexsym}
\usepackage{graphicx}
\usepackage{bm}
\usepackage{subfigure}

\begin{document}

\title{Three-Dimensional Topological Insulator in a Magnetic Field:
Chiral Side Surface States and Quantized Hall
Conductance}
\author{Yan-Yang Zhang}
 \affiliation{Department of Physics and Center of Computational and Theoretical Physics,
 The University of Hong Kong, Pokfulam Road, Hong Kong}
 \author{Xiang-Rong Wang}
 \affiliation{Department of Physics, The Hong Kong University of Science
and Technology, Clear Water Bay, Kowloon, Hong Kong and School of
Physics, Shandong University, Jinan, P. R. China}
 \author{X.C. Xie}
 \affiliation{International Center for Quantum Materials, Peking University, Beijing, China 100083 and Department of Physics,
 Oklahoma State University, Stillwater, OK 74078}
\date{\today}

\begin{abstract}
Low energy excitation of surface states of a three-dimensional
topological insulator (3DTI) can be described by Dirac fermions.
By using a tight-binding model, the transport properties of the
surface states in a uniform magnetic field is investigated.
It is found that chiral surface states parallel to the magnetic
field are responsible to the quantized Hall (QH) conductance
$(2n+1)\frac{e^2}{h}$ multiplied by the number of Dirac cones.
Due to the two-dimension (2D) nature of the surface states, the
robustness of the QH conductance against impurity scattering is
determined by the oddness and evenness of the Dirac cone number.
An experimental setup for transport measurement is proposed.
\end{abstract}


\maketitle

\section{I. Introduction}
Topological insulator (TI), a new quantum state in which the bulk is
an insulator and the surface is metallic, has attracted intensive
attention\cite{HasanReview} since its theoretical
predictions\cite{Fu2007,Fu2007b,Zhang2009} and its experimental
realizations\cite{Hsieh2009,Hsieh2009B,Roushan2009,YLChen2009} in
three dimensions. Unlike normal insulators, the boundary (edges or
surfaces) of a finite TI sample supports extended and current
carrying states in its bulk gap. In the presence of time reversal
symmetry, TIs can be classified by a $\mathbb{Z} _2$ invariant
integer $\nu$\cite{Fu2006,Moore2007,Qi2008} into a weak TI for even
number of the Dirac cones (called $\nu=1$) and a strong TI for odd
number of the Dirac cones (called $\nu=-1$). The edge/surface states
of a weak (strong) TI can (cannot) be destroyed by defect
scatterings\cite{Fu2007,Fu2007b}.

The surface states of a three-dimensional topological insulator
(3DTI) are described by two-dimensional (2D) Dirac cones centered
at the time-reversal invariance points in the Brillouin zone.
2D Dirac fermions with a single cone are predicted to have many
novel properties such as the Klein paradox, anti-localization,
etc.\cite{Katsnelson2006,Nomura2007,Shon1998,Ando1998}
In the quantum Hall effect (QHE) regime (strong magnetic field),
the Landau levels of a single cone Dirac fermion is unevenly
distributed
\begin{equation}
E_n\sim \sqrt{B|n|},\quad n=0,\pm1,\pm2,\cdots.\label{eq1}
\end{equation}
This distribution has been confirmed by several
experiments\cite{TZhang2009,Cheng2010,Hanaguri2010}. Another
interesting prediction of one Dirac cone fermion is the \emph{half}
(in the unit of $e^2/h$) quantized Hall (QH)
conductance\cite{Ando1998,Novoselov2006,Xiao2007} because one Dirac
cone carries a $\pi$ Berry phase. Although many properties of strong
3DTIs have been confirmed since its experimental realization in
several semiconductors, the half QH conductances has not been
observed in experiments yet, and this will be our focus.

In this paper, we investigate a 3DTI in a uniform magnetic field
within a well-known tight binding model. Our calculations yield
indeed the well-known Landau levels (Eq. (\ref{eq1})). However, the
issue of the half QH conductance is subtle. All surface states of a
finite 3DTI are fully connected with each
other\cite{Fu2007,Fu2007b,Lee09} and surface states living on the
sample surface may move from one side to another. In other words,
the well-defined Hall voltage measurement in usual 2D electron gases
(2DEG) is not so clear in the 3DTI cases since two surfaces cannot
be separated by the bulk of a finite 3DTI as it did in 2D QHE
systems. The current does not come from 1D channels, but from 2D
surface states. The structures of these side surface states are more
complicated than the edge states in 2DEG, therefore well-defined
quantum Hall plateaus $(2n+1)\frac{e^2}{h}$ can only be defined from
transverse current instead of the Hall voltage. Our tight-binding
calculations confirm the earlier predictions based on an effective
theory of Dirac fermion in curved 2D spaces\cite{Lee09}. The
robustness of the Hall conductance against impurities is also
studied. Our tight-binding model and their basic properties are
introduced in the next section. Numerical results and their
discussions are presented in section III, followed by the conclusion
section.

\section{II. Model}
A well-known tight-binding model of 3DTI\cite{Li10} is
\begin{align}
H_{0}(\mathbf{k})&=\epsilon_0(\mathbf{k})I_{4\times4}+
\sum_{a=1}^5 d_a(\mathbf{k})\Gamma^a \nonumber\\
d_a(\mathbf{k})&=(A_2\sin k_x,A_2\sin k_y,A_1 \sin
k_z,\mathcal{M}(k),0), \label{eq2}
\end{align}
where $\epsilon_0(\mathbf{k})=C+2D_1+4D_2-2D_1\cos k_z-2D_2(\cos
k_x+\cos k_y)$,$\mathcal{M}=M-2B_1-4B_2+2B_1\cos k_z+2B_2(\cos
k_x+\cos k_y)$ and $\Gamma^{1,2,3,4,5}=(\sigma_x\otimes
s_x,\sigma_x\otimes s_y,\sigma_y\otimes I_{2\times2},\sigma_z\otimes
I_{2\times2},\sigma_x \otimes s_z)$ in the basis of four states
$(|P1_z^+,\uparrow\rangle,|P1_z^+,\downarrow\rangle,|P2_z^-,
\uparrow\rangle,|P2_z^-,\downarrow\rangle)$. After inverse Fourier
transformation of equation (\ref{eq2}), the real space version of
this model on a cubic lattice can be written in the form
\begin{equation}
H_{0}=\sum_{is}\epsilon_{is}c^{\dagger}_{is}c_{is}
+\sum_{\langle ij\rangle \\
ss'}t^{ss'}_{ij}c^{\dagger}_{is}c_{js'}+\mathrm{H.c.} , \label{eq3}
\end{equation}
where $i,j$ are lattice site indices,
$s\in\{|P1_z^+,\uparrow\rangle,|P1_z^+,\downarrow\rangle,
|P2_z^-,\uparrow\rangle,|P2_z^-,\downarrow\rangle\}$
is the spin-orbital index. The first term in equation (\ref{eq3})
is the on-site energy, the second and third terms describe the
hoppings between nearest neighbor sites. The effect of non-magnetic
impurities can be included by adding a term
\begin{equation}
H_{I}=\sum_{is}V_{is}c^{\dagger}_{is}c_{is},  \label{eq4}
\end{equation}
where $V_{is}$ distributes randomly in the energy range of
$(-W/2,W/2)$. The magnetic field $\mathbf{B}$ is introduced
through the Peierls substitution of hopping
coefficients\cite{xrw1,Maciejko2009}
\begin{equation}
t^{ss'}_{ij}\rightarrow \exp(\frac{2\pi i}{\phi_0}\int_i^j
d\mathbf{l}\cdot \mathbf{A}) t^{ss'}_{ij}, \label{eq5}
\end{equation}
where $\phi_0=e/h$.

In the clean limit ($H_I=0$), this 3D model is fully gapped
in the energy range $(-M,M)$. The properties of surface
states within this gap are determined by the $\mathbb{Z}_2$
topological numbers related to the time reversal
polarizations\cite{Fu2006,Fu2007,Fu2007b}
\begin{equation}
\delta_i=\frac{\sqrt{\det[w(\Gamma_i)]}}{\mathrm{Pf}
[w(\Gamma_i)]}=-\mathrm{sgn}(d_4(\mathbf{k}=\Gamma_i))=\pm1,
\label{eq6}
\end{equation}
at 8 time-reversal invariant points in the first Brillouin
zone: $\Gamma_{1,2,3,4,5,6,7,8}=$ ($0,0,0$), ($\pi,0,0$),
($0,\pi,0$), ($0,0,\pi$), ($0,\pi,\pi$), ($\pi,0,\pi$),
($\pi,\pi,0$), ($\pi,\pi,\pi$), where $w_{mn}(\mathbf{k})\equiv
\langle u_{-\mathbf{k},m}|\Theta|u_{\mathbf{k},n}\rangle$.
If $\nu\equiv\prod_{i=1}^8\delta_i=-1$, the system is a strong
3DTI with odd numbers of the Dirac cones on \emph{each} surface.
Otherwise, the system ($\nu=1$) is a weak 3DTI with even numbers
of the Dirac cones on each surface. In terms of model parameters,
the system is in a strong (weak) TI phase when
$B_i/M>1$ ($B_i/M<-1$)\cite{Fu2007}.

The surface states around each Dirac cone is approximately
described by the Dirac fermion Hamiltonian (z-axis is normal
to the surface)\cite{Ando1998,Zhang2009,Li10,Shan2010}
\begin{equation}
H_{xy}=v_F(\sigma_x k_y-\sigma_y k_x),\label{eq7}
\end{equation}
with a linear dispersion relation
\begin{equation}
E_{xy}=\pm v_F\sqrt{k_x^2+k_y^2},\label{eq8}
\end{equation}
where $v_F=A_2\sqrt{1-(\frac{D1}{B1})^2}$ and $\hbar=1$ is adopted.

Before presenting calculations of tight binding models, let us
first look at effective model (Eq. (\ref{eq7})) in a field
$\mathbf{B}=(0,0,B)$ perpendicular to the surface. In the Landau
gauge $\mathbf{A}=(-By,0,0)$, $H_{xy}$ is\cite{Shon1998}
\begin{align}
H_{xy}&=v_F[\sigma_x(k_y-A_y)-\sigma_y(k_x-A_x)]\nonumber\\
&=v_F\left(
  \begin{array}{cc}
    0 & k_y+i(k_x+eBy) \\
    k_y-i(k_x+eBy) & 0 \\
  \end{array}\right)
,\label{eq9}
\end{align}
where $k_i=-i\partial_i$. The eigenvalues are Landau levels
\cite{Shon1998}
\begin{equation}
E_{xy}=\pm v_F\sqrt{2neB}, \quad n=0,1,2,\cdots \\ \label{eq10}
\end{equation}

However, for surface parallel to magnetic field, say, $x-z$ plane,
the Hamiltonian is
\begin{align}
H_{xz}&=\sigma_x(p_z-A_z)-\sigma_y(p_x-A_x)\\&= \left(
  \begin{array}{cc}
    0 & p_z+i(p_x+eBy) \\
    p_z-i(p_x+eBy) & 0 \\
  \end{array}\right).\label{eq11}
\end{align}
Since $eBz$ commutes with $k_x$ and $k_y$, the only difference
between Eq. (\ref{eq11}) and Eq. (\ref{eq7}) is a shift $-eBy$ of
the Dirac point in $k_x$ direction. The eigenvalue shows a shifted
Dirac cone
\begin{equation}
E_{xz}=\pm v_F\sqrt{(k_x+eBy)^2+k_z^2}.\label{eq12}
\end{equation}
We will see that states in the surfaces parallel to the magnetic
field play an important role in the QHE in 3DTI.
\begin{figure} [htbp]
\includegraphics*[bb=0 0 720 506,width=0.5\textwidth]{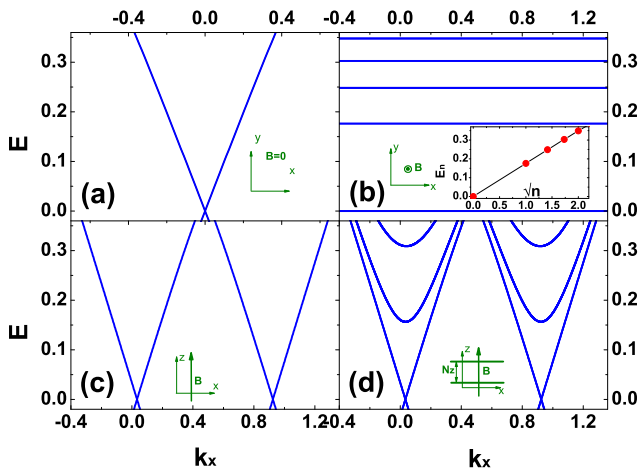}
\caption{(Color online) Dissipation relations of a strong 3DTI with
model parameters $A_1=A_2=1$, $B_1=B_2=1$, $C=0$, $D_1=D_2=0$ and
$M=0.4$. (a): $E(k_x,k_y=0)$ for a slab with geometry
$\infty\times\infty\times 40$ and without magnetic field. (b):
Similar to (a), but in a perpendicular magnetic field
$B_z=\frac{\phi_0}{400a^2}$. Inset of (b): the Landau levels $E_n$
versus $\sqrt{n}$. (c): $E(k_x,k_z=0)$ for a slab with geometry
$\infty\times60\times \infty$ in a parallel magnetic field
$B_z=\frac{\phi_0}{400a^2}$. (d) $E(k_x)$ for a bar of
$\infty\times60\times 40$ with periodic boundary condition in $z$
direction, in a magnetic field $B_z=\frac{\phi_0}{400a^2}$, in
contrast to its infinite-size version (c). } \label{F1}
\end{figure}

\section{III. Numerical Results}

To be specific, we consider Hamiltonian \eqref{eq3} on a cubic
lattice with size $N_x\times N_y\times N_z$. For a slab
$\infty\times \infty\times N_z$ of thickness $N_z$ in $z$-direction
and infinite in other two directions\cite{Fu2007b}, the surface
states of the TI can be displayed by plotting the dispersion
relation $E(k_x,k_y)$ inside the bulk gap so that this dispersion
relation must be from the surface states. Fig. \ref{F1} (a) is the
dispersion relation of a slab without a magnetic field. The Dirac
cone can be clearly seen. There are two degenerated Dirac cones
located on lower ($z=1$) and upper ($z=N_z$) surfaces, respectively.
The existence of such surface states roots on the topological
property of time reversal symmetric systems\cite{Fu2006,Fu2007}. The
magnetic field breaks time reversal symmetry and the fate of surface
states will be tested in the following tight-binding calculation. A
uniform magnetic field generally also breaks the lattice
translational symmetry\cite{xrw2} and results in the Hofstadter
butterfly spectrum\cite{Hofstadter}. If the magnetic flux through a
unit cell is a fraction of the flux quanta,
$Ba^2=\frac{p}{q}\phi_0$, where $p$ and $q$ are two prime numbers
and $a$ is the lattice constant, periodic structure is restored with
an enlarged and y-elongated unit cell of $q$-times of the original
one, $a'_x=a_x$ and $a'_y=qa_y$, then $k_x\in[-\pi/a,\pi/a$ and
$k_y\in[-\pi/(qa),-\pi/(qa)]$ are good quantum
numbers\cite{Thouless1982,Kohmoto1985}. Fig. \ref{F1} (b) is the
dispersion relation of the same slab as that for Fig. \ref{F1} (a)
in a perpendicular magnetic field $\mathbf{B}=(0,0,B)$. Instead of
linear dispersion relation in zero field, discrete Landau levels
$E_n$ appear. We have confirmed that there are an upper-lower
surface double degeneracy and all wavefunctions are strongly
confined inside (top/bottom) surfaces. Although the time-reversal
symmetry is broken, the magnetic field does not destroy the 2D
nature of the surface states. The dependence $E_n\sim \sqrt{n}$ is
also verified as shown in the inset of Fig. \ref{F1} (b). Such
Landau levels have already been observed in recent
experiments\cite{Cheng2010,Hanaguri2010}.

\begin{figure} [htbp]
\includegraphics*[width=0.5\textwidth]{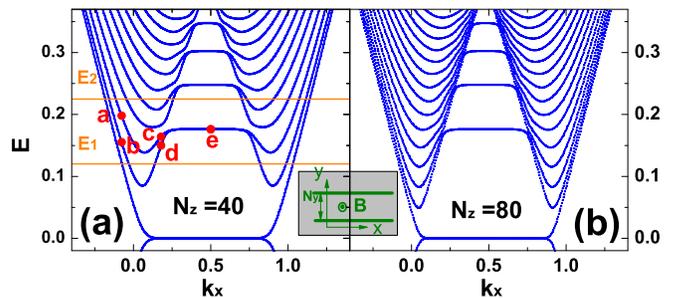}
\caption{(Color online) Dissipation relation $E(k_x)$ along the
$x-$direction of a bar geometry $\infty\times 60\times N_z$ in a
perpendicular magnetic field $B_z=\frac{\phi_0}{400a^2}$: (a)
$N_z=40$; and (b) $N_z=80$. Other model parameters are identical to
that for Fig. \ref{F1}.} \label{F10}
\end{figure}

We plot also the dispersion relation for the infinite $x-z$
surfaces in a parallel magnetic field $\mathbf{B}=(0,0,B)$
in Fig. \ref{F1} (c). Shifted Dirac cones corresponding to two
surfaces ($y=1$ and $y=N_y$) can be clearly seen, as predicted
in Eq. (\ref{eq12}). We will see that these side surface
states play an important role in the following discussion.
Fig. \ref{F1} (d) is similar to (c) when the infinite surface
is confined along $z$-direction with a periodic boundary
condition that eliminates possible edge states. Subband
structures appear seen due to the $z$-confinement.

With quantized Landau levels, one naturally expects quantum Hall
effects. In conventional 2DEGs, quantized Hall conductance
comes from the spatially separated counter-propagating edge
channel(s) on the two sides of a samples\cite{Halperin1982}.
The situation in 3D is subtle because a surface state may
cover all surfaces because they are fully
connected\cite{Fu2007,Fu2007b,Lee09}.
In other words, a surface state cannot be confined on only one
surface, resulting in more interesting with richer physics.

\begin{figure*} [htbp]
\includegraphics*[width=0.29\textwidth]{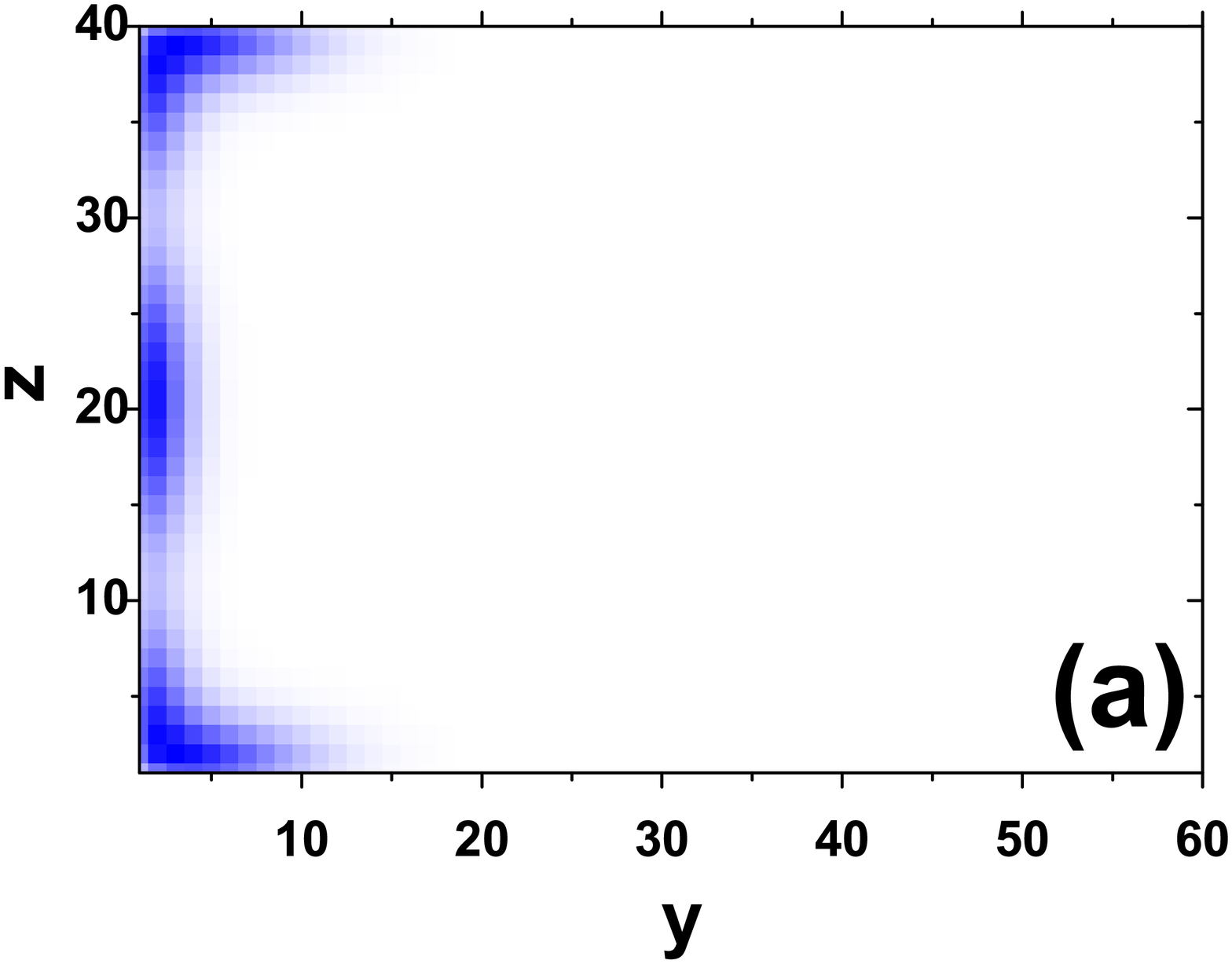}
\includegraphics*[width=0.3\textwidth]{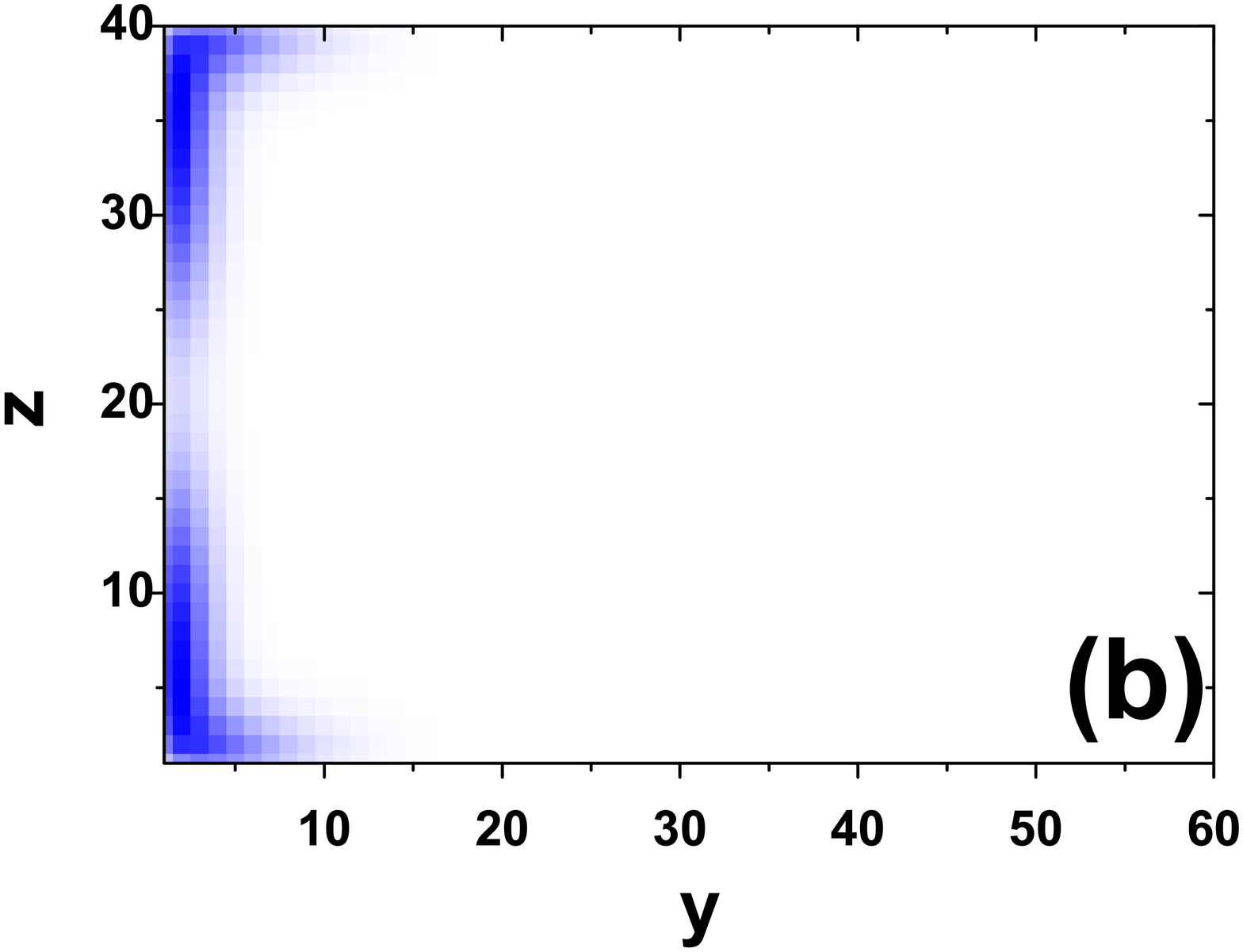}
\includegraphics*[width=0.3\textwidth]{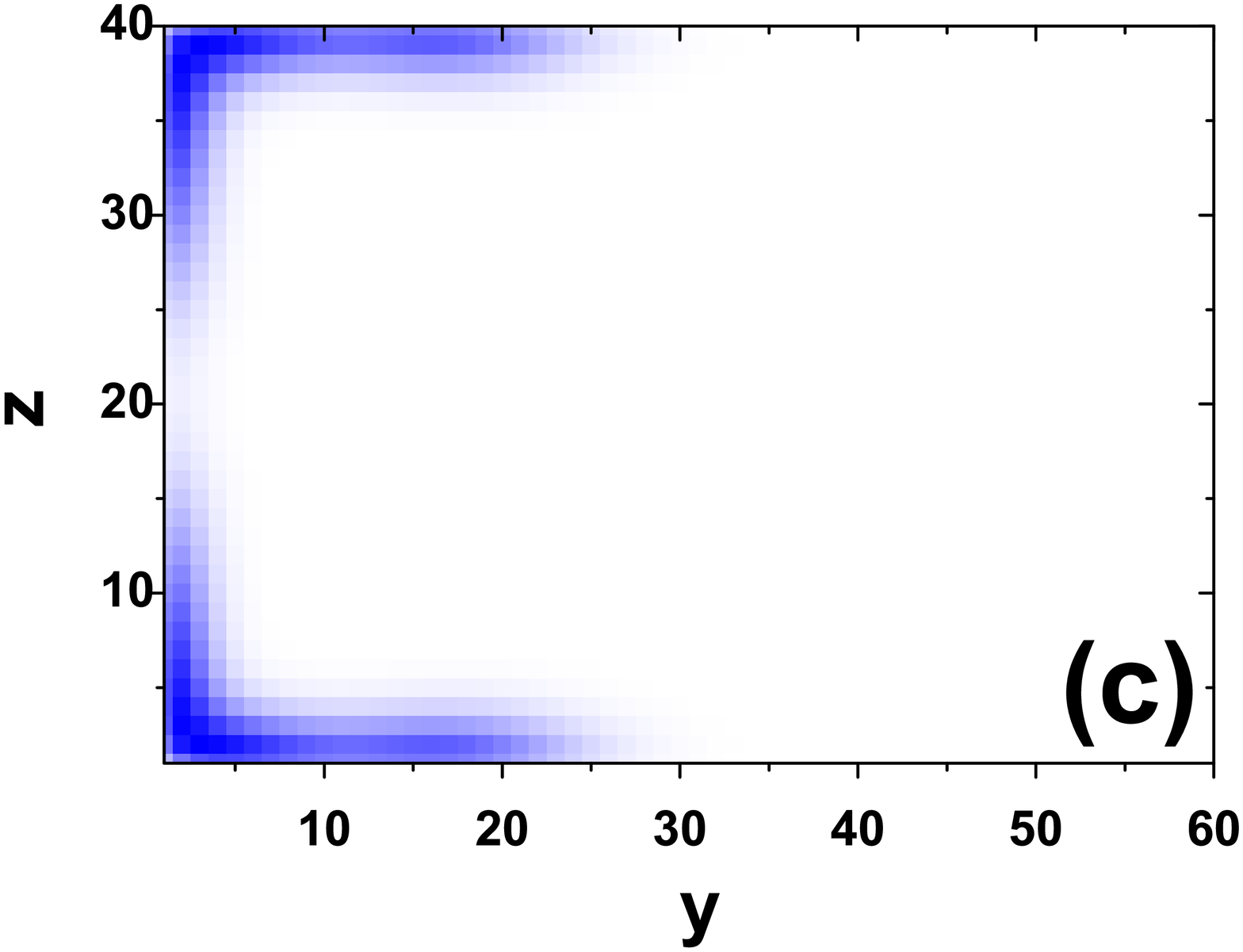}
\includegraphics*[width=0.3\textwidth]{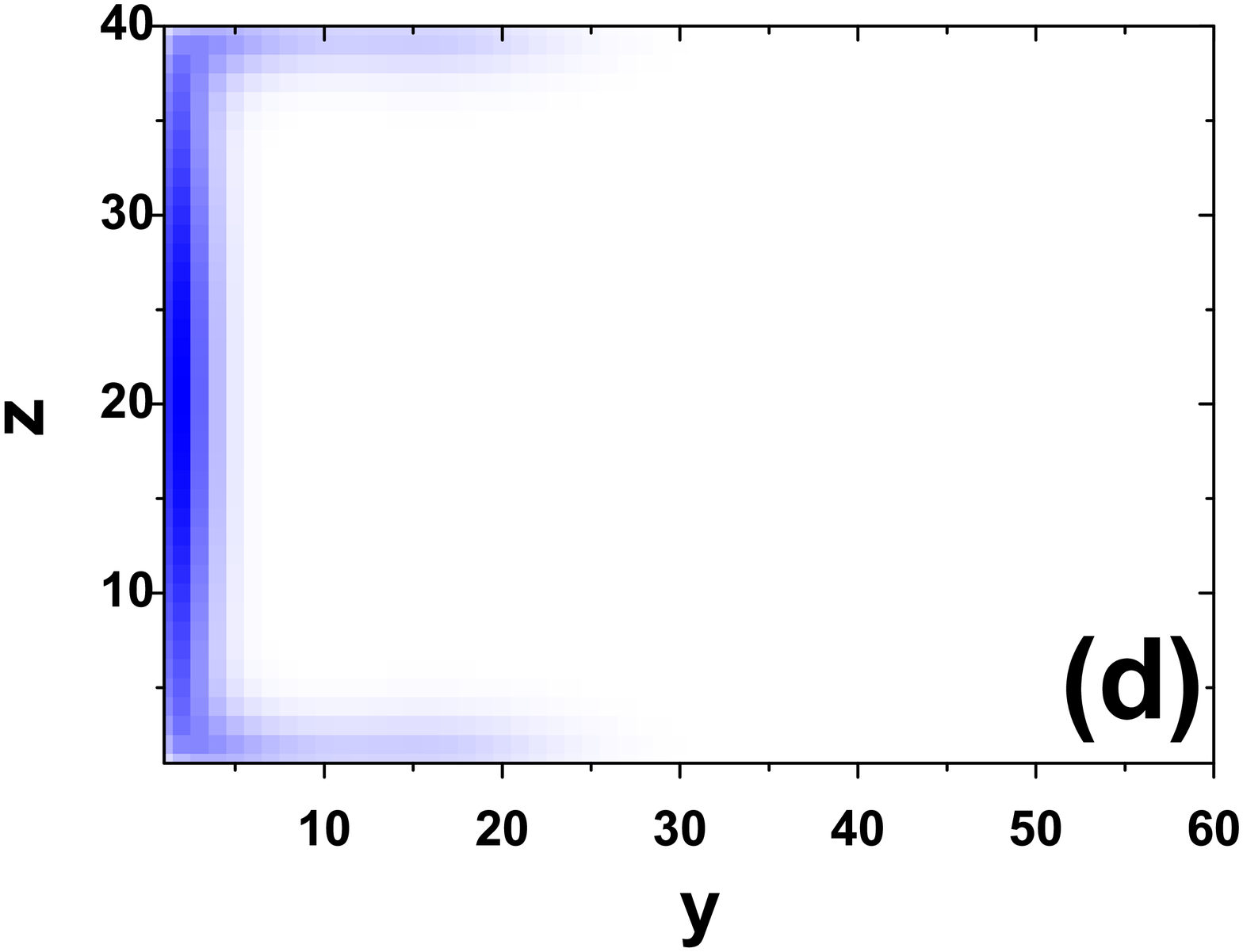}
\includegraphics*[width=0.3\textwidth]{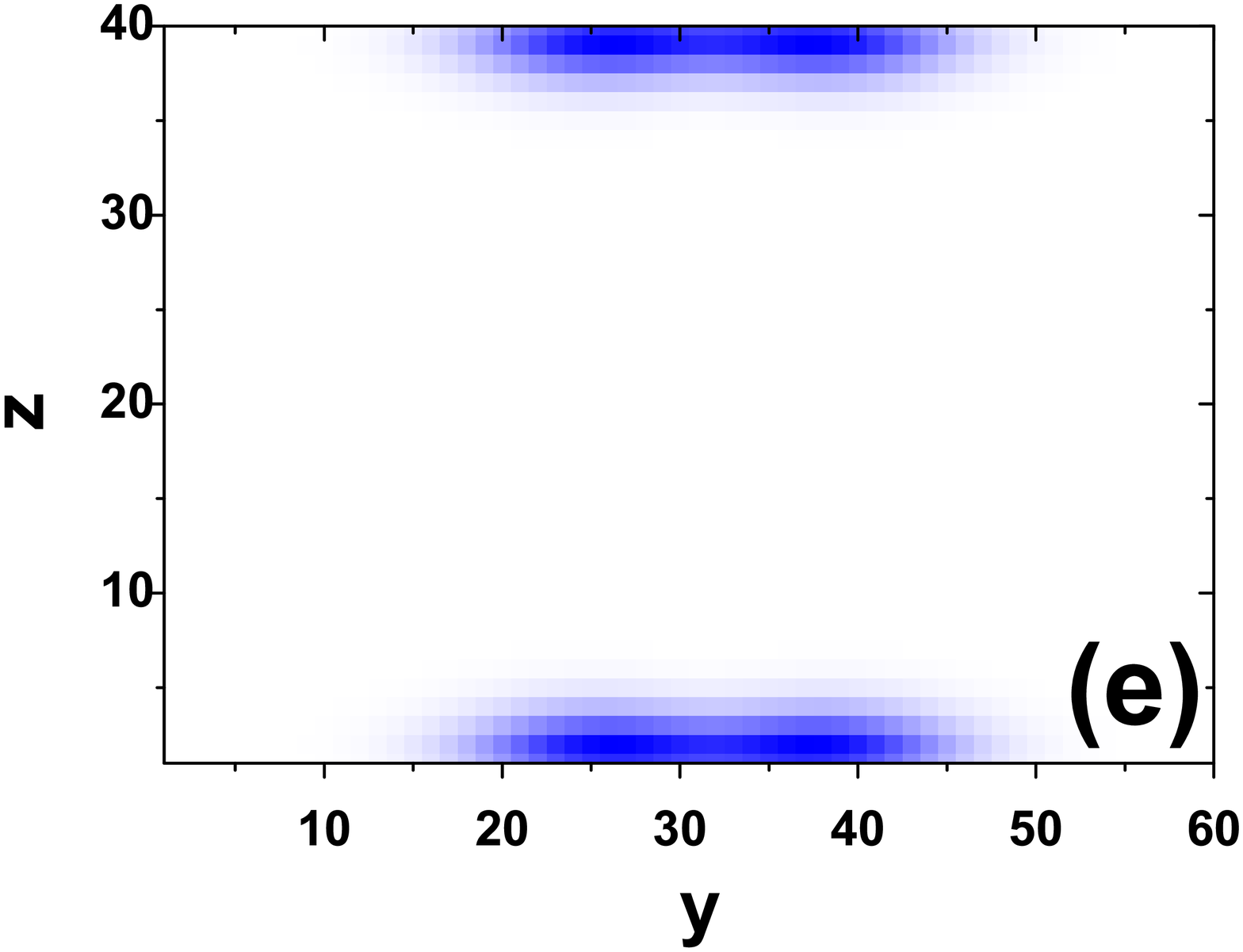}
\caption{(Color online) Spatial distributions of wavefunctions on
the $y-z$ section of the bar for the states marked as red dots in
Fig. \ref{F10} (a). The magnetic field is in $z$-direction. (a) to
(e) corresponds to state points a to e, respectively.} \label{F3}
\end{figure*}

\begin{figure} [htbp]
\includegraphics*[width=0.3\textwidth]{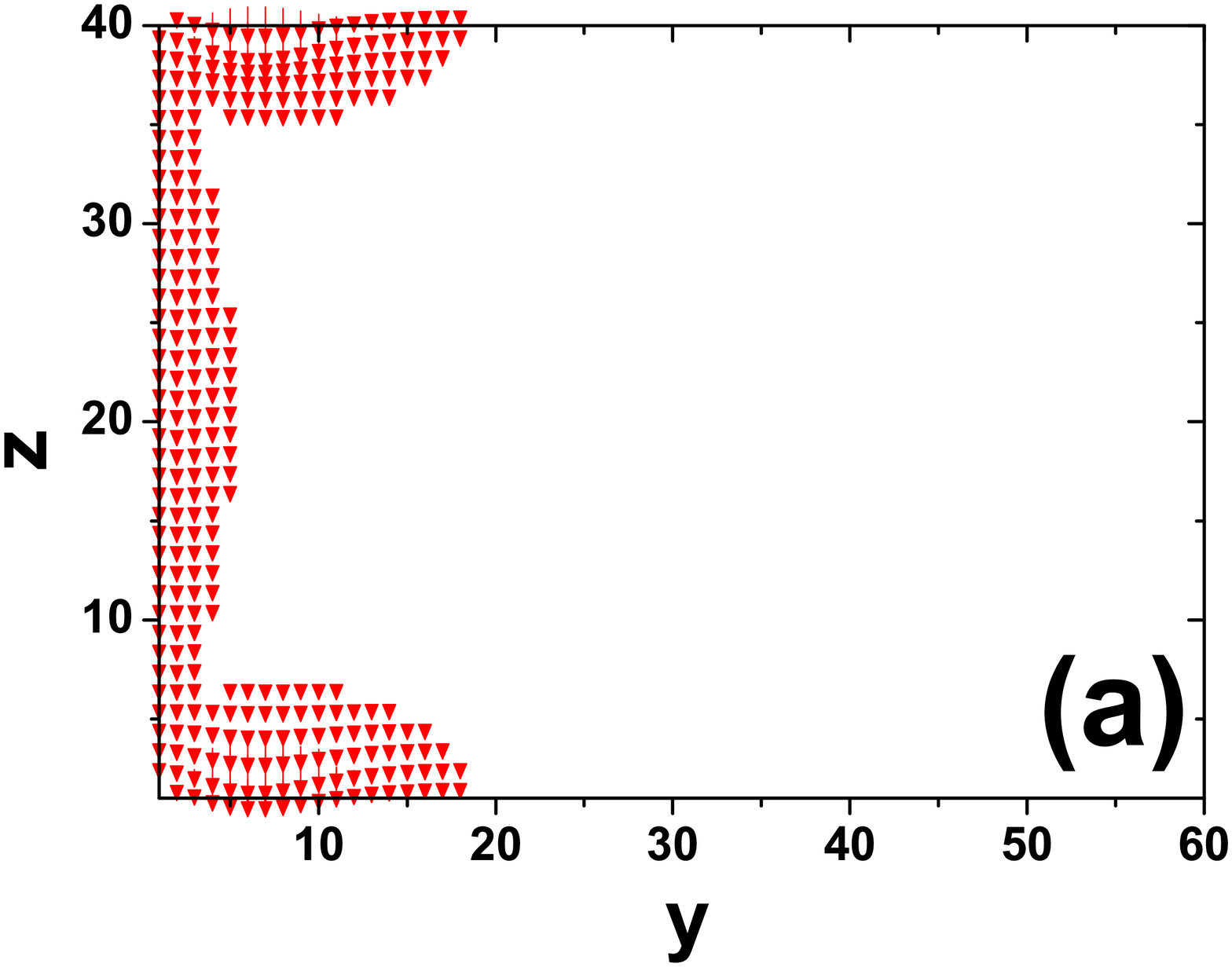}
\includegraphics*[width=0.3\textwidth]{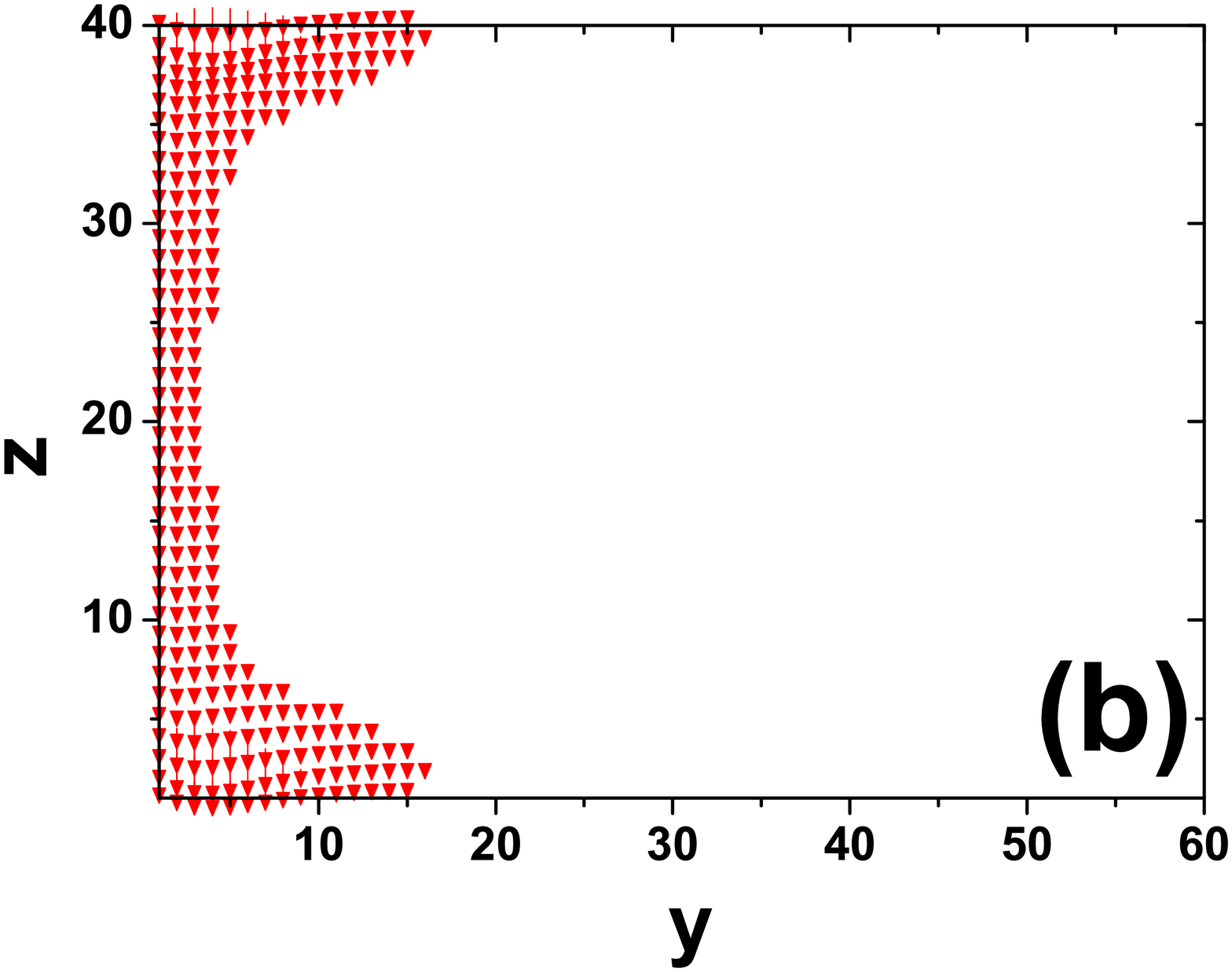}
\caption{(Color online) Spatial distributions of local spin $s_z$.
(a) and (b) correspond to state points a and b in Fig. \ref{F10},
respectively.} \label{Spin}
\end{figure}

To study possible current carrying surface states, we consider
a bar geometry of finite widths of $N_y$ and $N_z$ in both
$y-$ and $z-$ directions. The dispersion relation along
$x-$direction in a magnetic field is plotted in Fig. \ref{F10}.
The constant energy in the middle of the curves shows that the
Landau levels exist in the middle of the surface. The Landau
levels float up near the edges (junctions of two surfaces).
We have confirmed that the left-most ($k_x<0.1$) and
right-most ($k_x> 0.8$) parts of the curves correspond to
states distributed mostly on the left ($y= 1$) and the right
($y= N_y$) side surfaces, respectively (Fig. \ref{F3}).
These side surface states are spin-polarized due to spin-orbit
coupling\cite{Lee09}, as plotted in Fig. \ref{Spin}.

As illustrated in Fig. \ref{F3}(a) to (e), varying $k_x$ from either
side of the curves to the middle flat parts, the states move from
the side surfaces of $y=1$ ant $y=N_y$ (Fig. \ref{F1} (d)), to the
top and bottom surfaces of $z=1$ and $z=N_z$ (Fig. \ref{F1} (b))
continuously. This is possible because a particle can move from the
side surfaces to the top/bottom ones without passing through the
bulk. In other words, the surface states live essentially on a
closed surface of a finite 3DTI sample. One cannot have 1D edge
channels as in the 2D case. The discreteness of these states simply
originates from the finite thickness $N_z$ in $z$-direction, as
indicated in Fig. \ref{F10} (a) and (b). Increasing $N_z$ does not
affect the Landau levels, but generates more side surface-state
subbands. In the limit of large $N_z$, these subbands pack densely
and form two continuum cones, similar to the shifted Dirac cones in
parallel magnetic field in Fig. \ref{F1} (c).

\begin{figure} [htbp]
\includegraphics*[bb=0 0 731 573,width=0.4\textwidth]{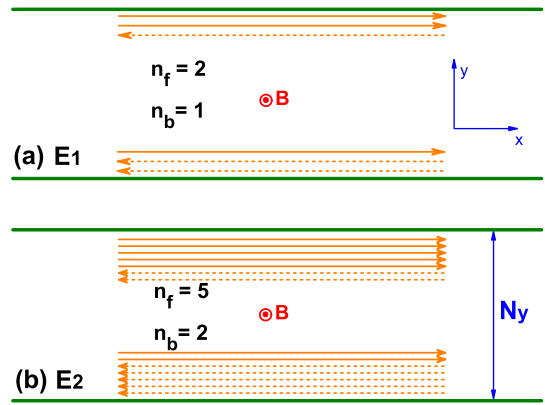}
\caption{(Color online) Schematic drawing (seen from above) of the
active side surface channels at the Fermi energies $E_1$ (a) and
$E_2$ (b) indicated by the orange lines in Fig. \ref{F10} (a). The
channel numbers of forward moving $n_{\mathrm{f}}$ and backward
moving $n_{\mathrm{b}}$ are indicated.} \label{FSchematic}
\end{figure}

\begin{figure} [htbp]
\includegraphics*[width=0.37\textwidth]{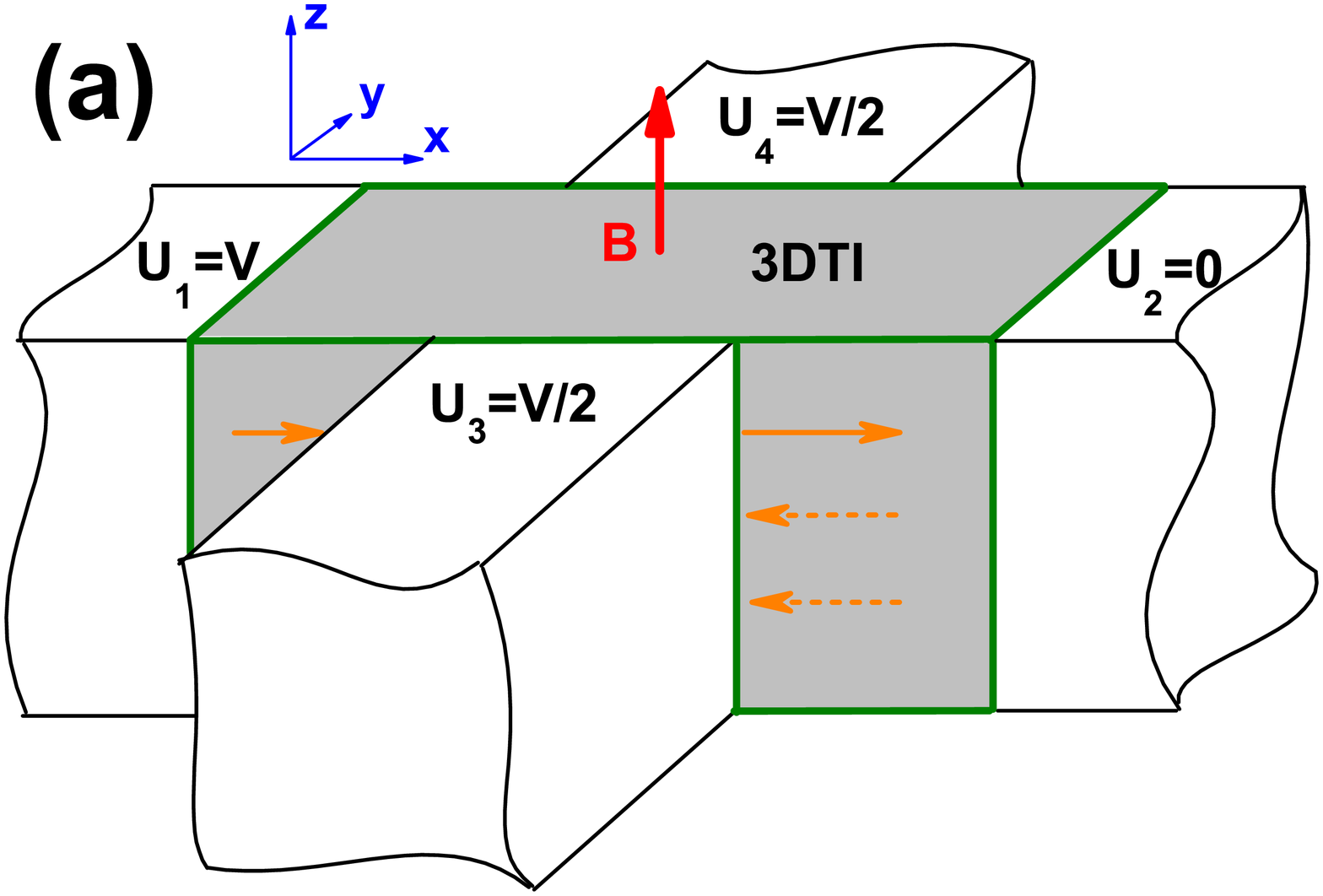}
\includegraphics*[bb=0 0 720 642,width=0.4\textwidth]{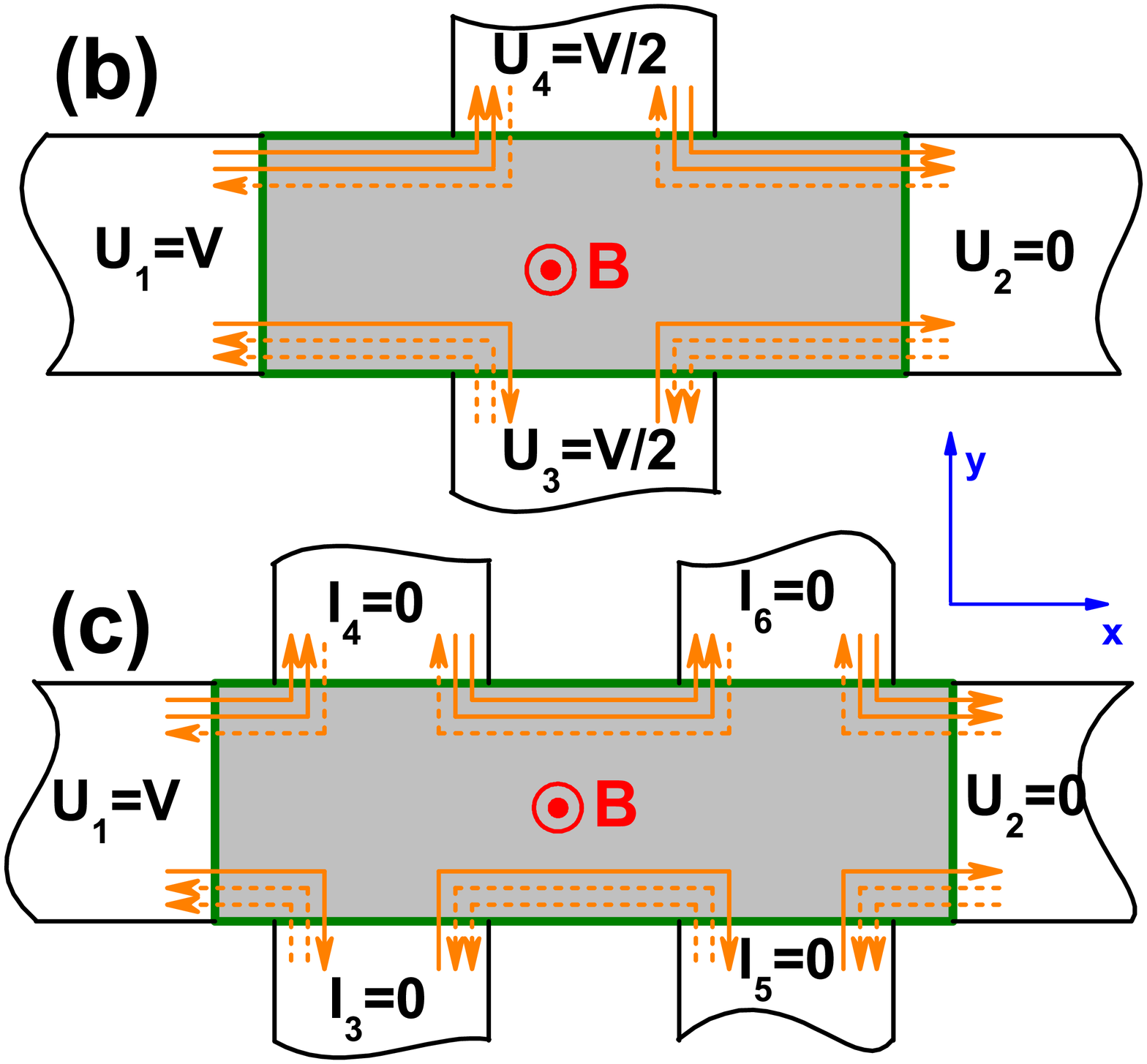}
\caption{(Color online) The schematic drawing of 3D multi-terminal
measurements of the TI (grey). The uniform magnetic field $B$ (red)
is in the $z-$ direction. The orange arrows represent the side
surface channels that carry currents. (a) The 3D schematic diagram
of a four-terminal measurement. A voltage $V$ is applied to the left
(1) and right (2) terminals. The potential of the front and back
terminals are $V/2$ and $V/2$, respectively. (b) The vertical view
of (a) from above. (c) The vertical view of a 3D six-terminal
measurement, which is not preferred in our model. }
\label{FMultiTerminal}
\end{figure}

Similar to edge states in conventional 2D systems, due to non-zero
dispersion, the side surface states can transport electrons in the
presence of voltage along $x$ direction. The effects of side surface
states have been noticed in a recent transport
measurement\cite{Brune2011}. However, the side surface states are
quite different from the edge states in 2DEG. For example, the
dependence of $E_n(k_x)$ is non-monotonic on each side. This brings
the coexistence of both forward and backward moving channels on each
side, as illustrated in Fig. \ref{FSchematic}. At each side and for
a given energy, the numbers of forward moving channels
$n_{\mathrm{f}}$ and backward moving $n_{\mathrm{b}}$ depend on
magnetic field $B$, as well as thickness $N_z$ of the sample.
However, their difference $n_{\mathrm{f}}-n_{\mathrm{b}}
=1,3,5,\cdots$ is universal between any definite pair of adjacent
Landau levels, as long as the dimension of the sample is not too
small to couple states on any opposite surfaces\cite{Shan2010}. A
general theorem guarantees the existence of current carrying
states\cite{xrw3}. The directions of these net currents on two sides
are reversed, thus they are chiral. Experimentally, such chiral
currents can be measured by a multi-terminal measurement illustrated
in Fig. \ref{FMultiTerminal}. In such setups, the
Landauer-B\"{u}ttiker formula is $I_j=\sum_iT_{j\leftarrow
i}\big[V_j-V_i\big]$, where $T_{j\leftarrow i}$ is the transmission
from lead $i$ to lead $j$, $V_i$ and $I_i$ are the voltage and the
current in the $i-$th lead respectively, satisfying Kirchoff's law
$\sum_i I_i=0$\cite{Datta}. In the clean limit, all the active
channels are perfectly conducting and the transmission
$T_{j\leftarrow i}$ can be read directly from the profile of the
active surface channels, as shown in Fig. \ref{FMultiTerminal}. For
the four-terminal setup illustrated in Fig. \ref{FMultiTerminal} (a)
and (b), straightforward calculations within this formalism give
\begin{align}
G_{xy}\equiv\frac{I_3-I_4}{V}=\frac{2I_3}{V}=-\frac{2I_4}{V}=(n_{\mathrm{f}}-n_{\mathrm{b}})\frac{e^2}{h},\label{eq13}\\
G_{xx}\equiv\frac{I_1-I_2}{V}=\frac{2I_1}{V}=-\frac{2I_2}{V}=(n_{\mathrm{f}}+n_{\mathrm{b}})\frac{e^2}{h}.\label{eq14}
\end{align}
This $G_{xy}$ is the Hall conductance in the present problem. This
is consistent with the previous calculation based on an effective
model in curved space\cite{Lee09}, but here understood within simple
band theories. Similar results hold for the case of surface states
with more than one Dirac cones and the only difference is the
conductance is multiplied by a factor of $N_{\mathrm{cone}}$, the
number of Dirac cones. However, for a 3D six-terminal setup, shown
in Fig. \ref{FMultiTerminal} (c) and generalized from the
traditional 2D Hall bar , similar calculations show that
\begin{eqnarray}
G_{xy}&=&\frac{V_3-V_4}{I_1}=\frac{n_{\mathrm{b}}-n_{\mathrm{f}}}{n_{\mathrm{b}}^2-n_{\mathrm{b}}n_{\mathrm{f}}+n_{\mathrm{f}}^2}\cdot\frac{e^2}{h},\\
G_{xx}&=&\frac{V_3-V_5}{I_1}=\frac{n_\mathrm{b}n_\mathrm{f}}{n_\mathrm{b}^3+n_\mathrm{f}^3}\cdot\frac{e^2}{h},
\end{eqnarray}
which is generally sample dependent and not quantized.

\begin{figure} [htbp]
\includegraphics*[width=0.5\textwidth]{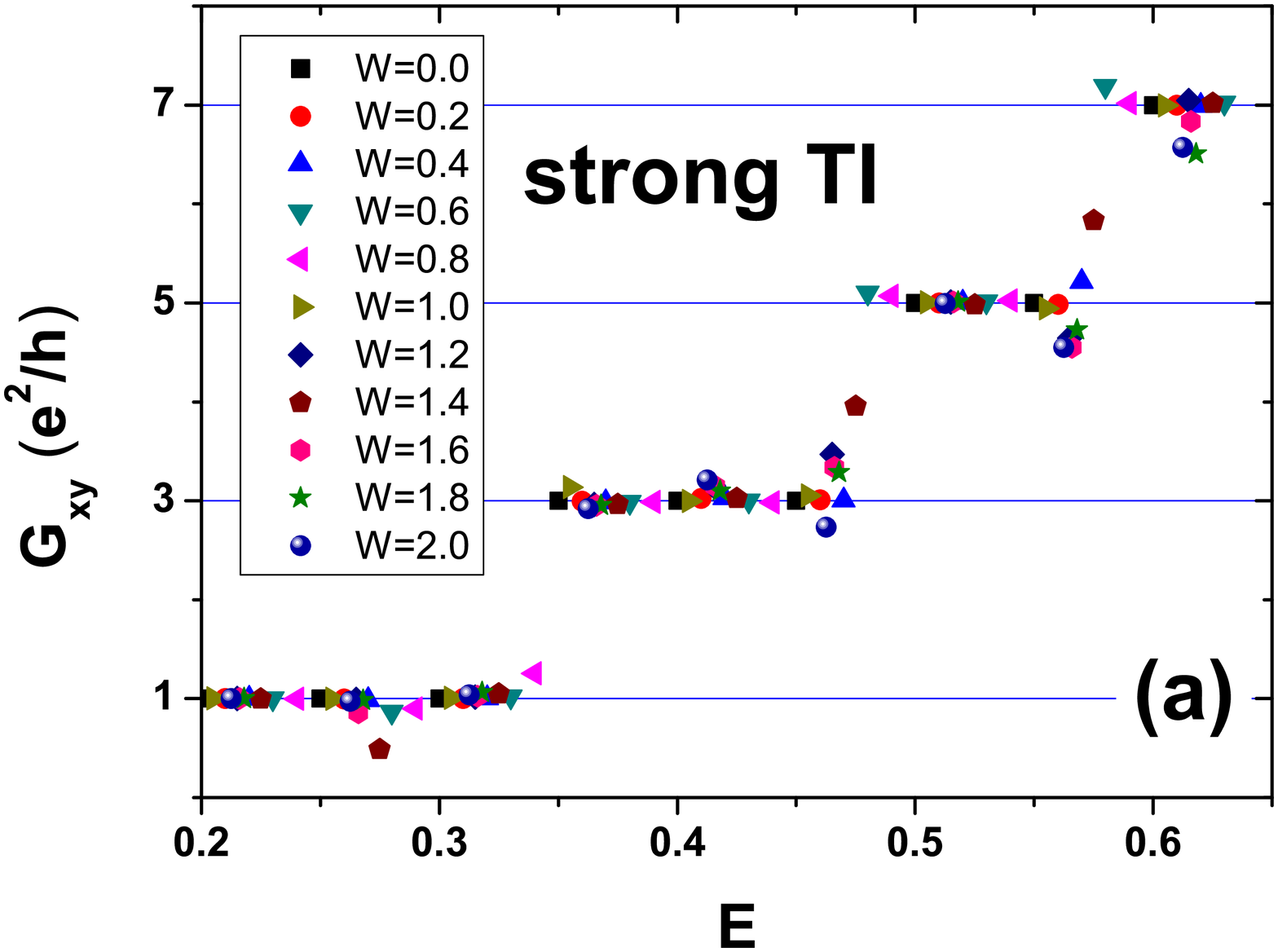}
\includegraphics*[width=0.5\textwidth]{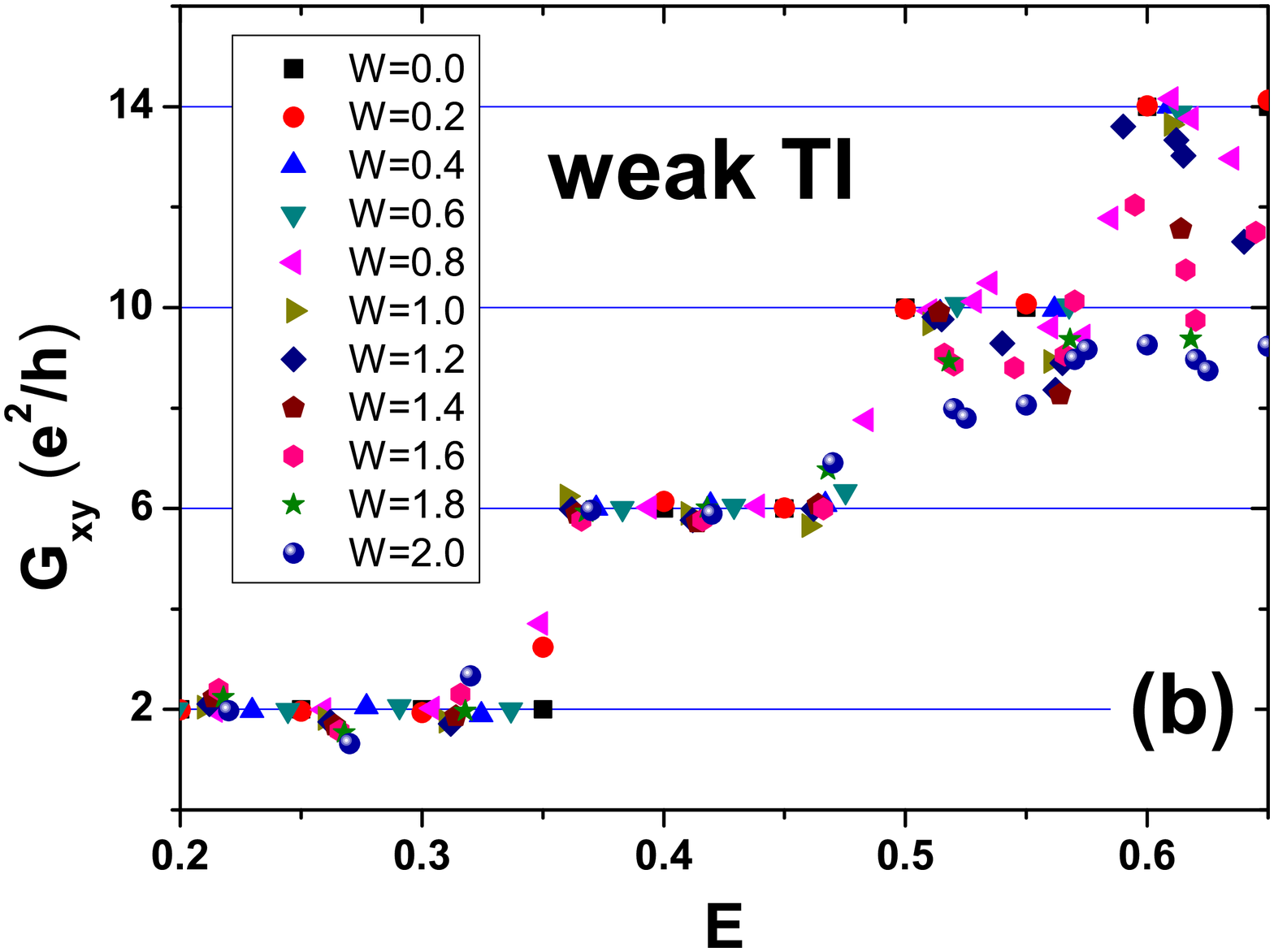}
\caption{(Color online) Hall conductance $G_{xy}$ as a function of
the Fermi energy $E$ for various disorder strength $W$. $G_{xy}$ for
a strong (a) and weak TI (b) is calculated from a four-terminal
setup as illustrated in Fig. \ref{FMultiTerminal}. The sample size
is $38\time36\times18$ and the model parameters are $A_1=A_2=1$,
$B_2=1$, $C=0$, $D_1=D_2=0$, $M=0.8$ and $B_1=\pm 1$ for strong
(weak) TI. } \label{QHE}
\end{figure}

One question arises: is the value of the Hall conductance $G_{xy}$
defined in Eq. (\ref{eq14}) robust against disorder\cite{Liu1996}?
In zero magnetic field, it is well-known that the surface states are
robust for odd $N_{\mathrm{cone}}$ (strong TI) and not for even
$N_{\mathrm{cone}}$ (weak TI). Since charge transport is dominated
by the side surface states at zero field, it is reasonable to expect
that the robustness of $G_{xy}$ is also determined by the evenness
and oddness of $N_{\mathrm{cone}}$. Specifically, for strong TI with
single-cone surface states in a magnetic field, as illustrated in
Fig. \ref{F10}, forward and backward channels coexist at the same
side, but they originate from a single Dirac cone and the
backscattering is small. These are confirmed by numerical
simulations of a four-terminal set-up as shown in Fig. \ref{QHE} (a)
for a strong 3DTI and in Fig. \ref{QHE} (b) for a weak 3DTI, by
using the standard method of non-equilibrium Green's
functions\cite{Datta,Guan2003}. In the weak TI case, the conductance
$G_{xy}$ is twice as large due to double cone degeneracy. Also, its
value is sensitive to disorder because of the inter-scattering
between the two cones. Since a four-terminal numerical simulation is
very resource consuming, in Fig. \ref{QHE} we use a smaller size
system. The physics remains the same as long as no direct coupling
between surface states on two opposite surfaces. Quantized Hall
plateaus predicted from above analysis, especially for the strong
TI, can be clearly seen. In the presence of disorder, the quantum
Hall plateaus of a strong TI (Fig. \ref {QHE} (a)) survive much
better than those of a weak TI (Fig. \ref{QHE} (b)).

\section{IV. Summary}
Before ending this paper, it should be emphasized that we did not
obtain the half conductance, or in a less strict sense, we saw the
sum of half conductance from upper and lower surfaces\cite{Lee09},
since the Dirac fermion live on a closed and curved surface of 3DTI.
By making an \emph{edge} channel on one surface, one inevitably
makes a new one on the opposite surface so that a Dirac fermion will
not terminate somewhere on a surface. The exact half Hall
conductance of one surface cannot directly be observed unless the
surface state can be effectively confined in one isolated plane by
some means. On the other hand, the quantized Hall conductance we
obtained from the 2D surface states should not be viewed as a
trivial application of Chern number theory of
2DEG\cite{Thouless1982,Kohmoto1985}, since the surface states live
on a closed 2D manifold embedded in a 3D space, which is
topologically different from the 2D plane of 2DEG.

In summary, we investigated the quantum Hall effect of a 3DTI in a
magnetic field. The integer Hall conductance is carried by side
surface states due to the non-separable nature of surface states
enclosing the 3DTI. The quantum Hall conductance reflects the
properties of side surface states that are parallel to the magnetic
field. The quantum Hall effect thus offers a transport measurement to
determine the topological property of a 3DTI: whether it is a weak or strong
TI, the number of Dirac cones, etc.

\section{Acknowledgements}
XCX is supported by NSF-China, MOST-China and
US-DOE-DE-FG-02-04ER46124. XRW is supported by HK CRF (No
HKU10/CRF/08-HKUST17/CRF/08) and RGC grants.

\end{document}